\documentclass[aps, twocolumn, showpacs]{revtex4}
\begin{document}
\title{Thermodynamics of Spacetime and Unimodular Relativity}
\author { S. C. Tiwari\\
Institute of Natural Philosophy\\
c/o 1 Kusum Kutir Mahamanapuri,Varanasi 221005, India}
\begin{abstract}
 The black hole entropy formula applied to local Rindler horizon at each spacetime point
has been used in the literature to derive the Einstein field equation as an equation 
of state of a thermodynamical system of spacetime. In the present paper we argue that
due to the key role of causal structure and discrete spacetime in this approach the 
natural framework is unimodular relativity rather than general relativity. It is shown that
the equation of state is trace free unimodular relativity field equation that uniquely
determines only the traceless stress tensor. Recent generalization to nonequilibrium
thermodynamics is shown to be equivalent to the conformally related spacetime metrics,
and energy-momentum conservation is satisfied without invoking entropy production.
We suggest that the cosmological constant should possess thermodynamical fluctuations,
and at a deeper level the metric may have statistical origin.
\end{abstract}
\maketitle

Black hole thermodynamics \cite{1} associates entropy with the area of event horizon
of a black hole. In recent years quantum gravity, though still elusive in a definite
form, has led some to believe that black hole thermodynamics would give new insights
into the nature of quantum space-time. Surprisingly one of the outstanding problems
of cosmology, namely the cosmological constant (CC) problem, seems linked with the entropy
of black hole calculated in terms of the accessible quantum states in some models; a lucid
nice exposition can be found in \cite{2}. In a remarkable different perspective \cite{3}
Jacobson argues that Einstein field equation could be viewed as an equation of state
of a thermodynamical system, and its canonical quantization may not make sense.
The most attractive aspect in this approach is that quantum space-time is envisaged
without quantum gravity. Obviously this deserves careful attention as an alternative to
various shades of quantum gravity theories. Further its philosophical
import is closer to the nature of space and time articulated in \cite{4}. An intriguing
part of the equation of state is that CC remains `as enigmatic as ever', and Jacobson
does not attempt to explain its significance. In a recent paper \cite{5} motivated by 
the effective field theory the recourse is taken to nonequilibrium thermodynamics, and
the generalized field equation (GFE) derived here does not have CC, but consistency 
with the energy-momentum conservation demands endowing shear viscosity to the horizon.

In this Letter we address the question asked by the authors in \cite{5} whether thermodynamic
derivation of Einstein field equation hints at "something deep". First, arguments are
presented to show that the natural setting for this approach is unimodular relativity
rather than general relativity. Next we modify the derivation of Jacobson considering the 
trace-free stress tensor for the heat flux, and obtain the field equation
without CC; let us call it UR field equation. That this result is not trivial becomes
clear with an important discovery: GFE could be recast in the form of UR field equation
using conformally related spacetime metric. This result is derived in a logically consistent
manner using the conformal transformation in the Raychaudhuri equation and the
heat-flux. It is also shown that covariant continuity of the equation of state
in general relativity framework  can be satisfied without introducing the entropy
production term. Thus a nontrivial approach beyond the one given in \cite{5} would be
required for nonequilibrium thermodynamics of spacetime.

Detailed review of \cite{3} is presented in the recent Letter \cite{5} therefore here we 
summarize the main ideas. The black hole formula, namely the proportionality between entropy
and the horizon area, is assumed to hold for all local Rindler horizons at each spacetime
point of the manifold M. Causal horizon at a point p is specified by a space-like
2-surface B, and the boundary of the past of B comprises of the congruences of null
geodesics. Assuming vanishing shear and expansion at p the past horizon of B is called
local Rindler horizon. The thermodynamical (Clausius) relation between entropy (S),
heat energy (Q), and temperature (T) given by $\delta Q=T dS$ is assumed. The energy
flux across the horizon is used for heat energy, and calculated in terms of the boost energy of matter: define an approximate boost Killing vector field future pointing on the causal 
horizon $\chi^\mu$ related with the horizon tangent vector $k^\mu$ and affine 
parameter $\lambda$ by $\chi^\mu=-\lambda k^\mu$. The heat flux is given by
\begin{equation}
\delta Q=\int T_{\mu\nu} \chi^\mu d\Sigma^\nu
\end{equation}
The integral is taken over a small region of pencil of generators of the inside past horizon terminating at p.  If area element is $dA$ then $d\Sigma ^\nu=k^\nu d\lambda dA$, and Eq.(1)
becomes
\begin{equation}
\delta Q=-\int T_{\mu\nu} k^\mu k^\nu \lambda d\lambda dA
\end{equation}
Change in the horizon area is given in terms of the expansion of the congruence of null geodesics generating the horizon $\delta A=\int \theta d\lambda dA$. The quantum vacuum in flat spacetime for the generator of Lorentz boosts could be treated as a Gibbs ensemble with temperature $T=\hbar /2\pi$. Note that a uniformly accelerated observer behaves as if immersed in a thermal bath at the Unruh temperature \cite{6} $T_U=acceleration \times T$. Now the equation of geodesic deviation for null geodesic congruence is given by the Raychaudhuri equation
\begin{equation}
\frac {d\theta}{d\lambda}=-\frac{\theta^2}{2}-\sigma_{\mu\nu}\sigma^{\mu\nu}-R_{\mu\nu}k^\mu k^\nu
\end{equation}
Assuming vanishing shear and neglecting $\theta ^2$ term we get the solution
\begin{equation}
\theta=-\lambda R_{\mu\nu} k^\mu k^\nu
\end{equation}
Assuming universal entropy density $\alpha$ per unit horizon area it is straightforward to
calculate the entropy change
\begin{equation}
\delta S=-\alpha \int R_{\mu\nu} k^\mu k^\nu \lambda d\lambda dA
\end{equation}
The condition that the Clausius relation is satisfied for all null vectors $k^\mu$ gives
\begin{equation}
R_{\mu\nu} + \Phi g_{\mu\nu}=(2\pi /\hbar\alpha) T_{\mu\nu}
\end{equation}
The unknown function $\Phi$ is determined using the covariant divergence law for the stress tensor and contracted Bianchi identity; Einstein equation with a cosmological constant $\Lambda$ is obtained. Here Newton's gravitational constant is identified as $G=1/4\hbar\alpha$.

Let us recall that causal structure of spacetime specifies past-future relationship of events and leads to the conformal geometry of the manifold M. Secondly the elementary entities akin
to ideal gas molecules could be imagined as some kind of granules of spacetime forming
the Gibbs ensemble. These basic considerations indicate unimodular relativity framework
rather than general relativity for Jacobson's approach. In an interesting paper Anderson
and Finkelstein \cite{7} proposed cellular spacetime in analogy with the phase space
in quantum theory. A measure manifold with a fundamental measure $\mu (x)$ restricts general relativity to unimodular relativity such that the metric determinant is given by
\begin{equation}
\sqrt {- g} d^4 x=\mu (x) d^4 x
\end{equation}
In this theory the field variable is a conformal metric tensor. The measure of spacetime is counting the number of modules in a fixed volume element of a discrete structure. The differentiable manifold M is a smoothed out macroscopic spacetime; CC can be shown to arise as an integration constant. Since metric tensor is not specified within the approach of \cite{3} it is not possible to
invoke conformal transformation directly. However we have shown that the covariant divergence
law of stress tensor is ambiguous \cite{8} and can be modified, and the trace free field 
equation is natural in unimodular relativity. Thus we make an ansatz: instead of stress tensor
the trace free stress tensor
\begin{equation}
S_{\mu\nu}= T_{\mu\nu} -\frac {1}{4} g_{\mu\nu} T^{tr}
\end{equation}
should appear in the heat flux $\delta Q$, Eq.(1), and that the energy-momentum conservation
used by Jacobson could be dispensed with. As a result tracing Eq.(6) we determine
$\Phi =-R/4$, and the equation of state is the UR field equation
\begin{equation}
R_{\mu\nu} - \frac {1}{4} g_{\mu\nu} R=\frac {2\pi}{\hbar\alpha} S_{\mu\nu}
\end{equation}
As is well known one can enforce energy-momentum conservation and transform this equation
into the Einstein field equation with CC, however we will return to CC after discussing
the recent consideration of nonequilibrium thermodynamics in \cite{5}. To incorporate
higher curvature terms in the Einstein field equation indicated by effective field theories
authors suggest a curvature correction to entropy, and assume entropy density to be 
$\alpha f(R)$. The entropy change is calculated similar to Eq.(5) given by
\begin{equation}
\delta S=\alpha \int (\theta f+\dot{f}) d\lambda dA
\end{equation}
with overdot denoting derivative with respect to $\lambda$. Einstein equation is replaced by
the GFE
\begin{equation}
f R_{\mu\nu} - f_{;\mu\nu}+\frac {3}{2} f^{-1} f_{,\mu} f_{,\nu} +\Psi g_{\mu\nu}=\frac {2\pi}
{\hbar\alpha} T_{\mu\nu}
\end{equation}
Here $\Psi$ is undetermined function. Authors argue that this equation is inconsistent with
the energy-momentum conservation, and propose entropy production for a nonequilibrium
system to make it consistent with the conservation law. The sole aim of entropy production is
to introduce a term that exactly balances the leftout term in the covariant divergence of the GFE.  Instead of this we investigate GFE in the unimodular relativity framework.

Following our ansatz we assume trace free stress tensor on the RHS of GFE and determine the function $\Psi$ by taking trace of Eq.(11) to obtain
\begin{equation}
-4\Psi = fR-f^{:\mu}_{~:\mu}+\frac {3}{2} f^{-1} f_{,\mu}f^{,\mu}
\end{equation}
Substitution of $\Psi$ in Eq.(11) finally gives the modified GFE using the notation
$\bar{R}_{\mu\nu} =R_{\mu\nu}-\frac{1}{4} g_{\mu\nu} R$
\begin{equation}
f\bar{R}_{\mu\nu}-f_{;\mu\nu}+\frac {3}{2} f^{-1} f_{,\mu}f_{,\nu}
+\frac {1}{4}g_{\mu\nu}(f^{:\alpha}_{~:\alpha}-\frac{3}{2} f^{-1} f_{,\alpha}f^{,\alpha})=\frac {2\pi}{\hbar\alpha} S_{\mu\nu}
\end{equation}

To interpret the modified GFE let us consider conformal metric
\begin{equation}
\tilde{g} _{\mu\nu}=f g_{\mu\nu}
\end{equation}
for which the Ricci and scalar curvatures are given by
\begin{equation}
\tilde{R}_{\mu\nu}=R_{\mu\nu}+f^{-1}[f_{;\mu\nu}-\frac {3}{2} f^{-1} f_{,\mu} f_{,\nu}
+\frac {1}{2} g_{\mu\nu} f^{:\alpha}_{~:\alpha}]
\end{equation}
\begin{equation}
\tilde{R}=f^{-1}[R+3f^{-1} f^{:\alpha}_{~:\alpha}-\frac {3}{2} f^{-2} f_{,\alpha} f^{,\alpha}]
\end{equation}
In deriving Eq.(13) no consideration was given to the underlying metric tensor, therefore,
it is possible to assume that instead of $\bar{R}_{\mu\nu}$ we have curvatures in the 
conformal frame; let us replace it by $(\tilde{R}_{\mu\nu} -\frac {1}{4}f g_{\mu\nu} \tilde{R})$.
Substitution of relations (15) and (16) in the resulting equation shows that the terms
containing the derivatives of f cancel out on the LHS, and we get
\begin{equation}
f \bar{R}_{\mu\nu}=\frac {2\pi}{\hbar\alpha} S_{\mu\nu}
\end{equation}
If stress tensor is multiplied by f then this equation is identically the UR field equation (9).
Recent discussion shows \cite{9} that it is only the trace free stress tensor that is uniquely defined in unimodular relativity; thus invoking trace free stress tensor to implement
unimodular relativity is justified.

In the preceding derivation, suggested replacement of the curvature tensors seems arbitrary. The reason for this apparent arbitrariness lies in the fact that in Jacobson's approach conformal
transformation is not an issue, and Raychaudhuri equation remains unaltered for the field
equations with higher curvature terms \cite{10}. Thus entropy density can be assumed to be
$\alpha f(R)$ leading to the GFE. In the present conformal perspective the entropy change
is interpreted in terms of a conformal transformation since entropy is related with a 
geometrical quantity (area). This implies that conformal transformation must be employed
consistently in the Raychaudhuri equation and the heat flux: the factor $f(R)$ should
appear in the integrand of Eq.(2) due to area element, and conformal Ricci tensor should
occur in Eq.(3). Following the steps that led to GFE we obtain the equation in general relativity framework  
\begin{equation}
f\tilde{R}_{\mu\nu} - f_{;\mu\nu}+\frac {3}{2} f^{-1} f_{,\mu} f_{,\nu} +\Psi \tilde{g}_{\mu\nu}=\frac {2\pi}{\hbar\alpha} f T_{\mu\nu}
\end{equation}
Using expressions (14) and (15) this equation reduces to
\begin{equation}
R_{\mu\nu} + \frac {1}{2} f^{-1} f^{:\alpha}_{~:\alpha} g_{\mu\nu} +\Psi g_{\mu\nu}=\frac {2\pi}{\hbar\alpha} T_{\mu\nu}
\end{equation}
It is easily verified that in unimodular relativity using trace free stress tensor we finally
get the UR field equation (9) as claimed; moreover there is no ambiguity of the factor of
f as in Eq.(17).

An important outcome of this derivation is that in general relativity framework, Eq.(19) is consistent with the energy-momentum conservation. Obviously there is no need to invoke entropy production. Further the Einstein field equation with CC is obtained: take covariant divergence of
Eq.(19) and use contracted Bianchi identity and covariant divergence law for stress tensor. 
However in the spirit of Jacobson's thermodynamic
approach CC should have thermodynamic significance. In this context we note the correspondence
between the causal set theory \cite{11} and unimodular relativity. In unimodular relativity
four volume $\tau$ and $\Lambda$ can be treated as canonically conjugate variables similar to
energy and time in quantum mechanics. Causal set
theory postulates discrete elements of spacetime, and the number N of such elements
is related with the volume $\tau$. Since N is subjected to Poissonian fluctuations
we have as a consequence a fluctuating CC. In an interesting paper \cite{12} this idea is 
explored to explain the dark energy in terms of a fluctuating CC. The assumption
of causal set theory as an underlying theory for unimodular relativity leads to 
quantum fluctuations in CC. However following Jacobson's thermodynamic approach in
unimodular framework the fluctuations in CC have origin in thermodynamics. It seems this
interpretation could be useful for investigating the dark energy problem in terms of a
fluctuating cosmological constant $\Lambda$.  

To summarize, there are two principal results of our work: A) The basic ingredients
of null geodesic congruence and discrete spacetime in this thermodynamic approach
indicate unimodular relativity framework, and the equation of state is the
trace free unimodular field equation. And, B) Nonequilibrium thermodynamics a la black
hole entropy formula with curvature correction term has been shown
to be equivalent to conformally related spacetime. Einstein equation with CC is once again
the equation of state. These considerations suggest thermodynamic significance to CC.

Result B has far reaching significance on the question of nonequilibrium thermodynamics of spacetime as a nontrivial approach becomes imperative; we make some remarks on the possible new avenues in the following.

1. Statistical origin of the metric- Jacobson \cite{3} recognises the
importance of fluctuations as temperature and entropy are not well defined for a 
nonequilibrium system and notes that, 'We speculate that out of equilibrium vacuum
fluctuations would entail an ill-defined spacetime metric'. In our approach we gain
better insight keeping in mind the implications of causal set theory and unimodular relativity.
Does the infinitesimal line element in general relativity represent a measure of
correlation between the discrete elements? Recall that in the statistical theory of fluctuations
\cite{13}, there is defined a mean value
\begin{equation}
\overline{(n_1-\bar{n}_1)(n_2-\bar{n}_2)}=\overline{n_1n_2} -\bar{n}^2
\end{equation}
where $n_1$ and $n_2$ are the particle number densities for a monoatomic substance
at two different points in space, and $\bar{n}$ is the mean density which for homogeneity
implies that $\bar{n}_1=\bar{n}_2=\bar{n}$. If there are no correlations the mean value defined
by Eq.(20) vanishes. Recall that one can envisage
the ever increasing number N of causal set elements as a kind of time variable \cite{12}.
If the line element is assumed to be a two-point correlation similar to Eq. (20) then the fourth coordinate
is the mean number disguised as time. The fundamental question related with the Lorentz
signature of the spacetime metric thus finds statistical explanation. Though challenging
yet immensely fruitful problem would be to link entropy increase with physical time 
following this idea.

2. Nonequilibrium thermodynamics- Recent paper \cite{5} has raised the question whether
nonequilibrium thermodynamical approach could have another justification. In Jacobson's
paper \cite{3} equivalence principle was used for uniform acceleration in the
neighbourhood of each spacetime point in order to apply equilibrium thermodynamics.
Reflecting on the foundations of classical general relativity Bondi in a cogent
argument \cite{14} shows that physical equivalence of accelerated frames does not hold.
The gravitation is observable because acceleration varies at each spacetime point.
The universal observable is the relative acceleration of neighbouring particles and
simplest choice is that it is proportional to the Riemann-Christoeffel tensor
$R_{\mu\nu\lambda\sigma}$. The Unruh temperature, on the other hand. is defined
for uniform acceleration. Logically we expect nonuniform temperature or a temperature
field to account for the observable gravitation. Obviously it would be a nonequilibrium
thermodynamical system. It is not clear in what way these considerations could be 
incorporated in Jacobson's approach.  

An alternative discussed in \cite{5} is to include nonvanishing shear term in the Raychaudhuri equation (3). In a different perspective heuristic analogy for entropy production
based on the form of entropy change expressed in terms of Ricci tensor in Eq.(5) and the
Ricci flow equation as a gradient flow proposed by Parelman \cite{15} offers another
possible approach. In section 5 of \cite{15} statistical analogy for a closed manifold with probability measure leads to the identification of (minus) action functional of the Riemannian manifold with entropy. It seems the evolution equation for Ricci flow
\begin{equation}
\frac{d g_{\mu\nu} (\tau)}{d\tau} = -2 R_{\mu\nu}
\end{equation}
as a gradient flow given by
\begin{equation}
\frac {d g_{\mu\nu}}{d\tau} = -(2 R_{\mu\nu} + u_{;\mu\nu})
\end{equation}
indicates that Eq. (5) could be generalized to entropy flow equation with an additional term
$f_{;\mu\nu}$ in the integrand. Note that in Parelman's work the geometry is Riemannian
and $\tau$ is the evolution parameter; u is a scalar function. Definition of entropy
is also different- in Jacobson's approach we are not using action functional. We leave the question of energy-momentum conservation and logically consistent formulation of nonequilibrium
thermodynamics of spacetime based on Ricci flow and Jacobson's approach for future.

I thank Prof. T. Jacobson for helpful correspondence, and Dr. W. Graf for drawing my attention to \cite{15}. Library facility of Banaras Hindu University is acknowledged.

\end{document}